# Coherent Averaging: an Alternative to the Average Hamiltonian Theory

A. K. Khitrin[1], Jiadi Xu[2], and Ayyalusamy Ramamoorthy[2]

[1]*Department of Chemistry and Biochemistry, Kent State University, Kent, Ohio 44240-0001, USA*
[2]*Biophysics and Department of Chemistry, University of Michigan, Ann Arbor, Michigan 48109-1055, USA*

Line-narrowing by periodic modulation of nuclear spin interaction Hamiltonians is the central element of various experimental techniques in NMR spectroscopy. In this study, we present a theoretical formulation of coherent averaging to calculate the heights of narrowed spectral peaks. This concept is experimentally demonstrated using proton spectra of solids obtained under fast magic-angle spinning.

Anisotropic interactions in many-body systems can be averaged by the application of external fields or by mechanical motion of the sample. This type of averaging can be called coherent to distinguish it from averaging by stochastic molecular motions. Due to low frequencies and feasibility of experimental implementations, such line-narrowing techniques have mainly been developed and utilized in nuclear magnetic resonance (NMR) spectroscopy. Many theoretical approaches to coherent averaging, although quite general, have also been formulated in the context of solid-state NMR spectroscopy. To name some, they are the average Hamiltonian theory (AHT) [1,2], canonical transformation technique [3], method of non-equilibrium statistical operator [4], and Floquet theory [5]. Equivalence of different effective Hamiltonians has been discussed in [6]. In a similar way, the average Liouvillian theory (ALT) [7-9] can also be developed. Here we will use the term AHT very broadly to name a class of theories which derive results as expansions in powers of the interaction. Usually, only few terms of such expansions can be calculated, and very frequently a non-vanishing term of the lowest order is linked to experimental observations. For a Hamiltonian $H(t)$ in an interaction frame with zero simple time average $<H(t)> = 0$, higher-order terms responsible for the residual line width appear when $H(t)$ does not commute with itself at different time intervals.

From a practical point of view, the AHT approach has several weak points. Formal requirement for neglecting the higher-order terms is $\Omega >> |H|$, where $\Omega$ is the modulation frequency. This requirement is not achievable in experimental applications, like homo- or hetero-nuclear spin decoupling, or magic-angle spinning (MAS) of the sample in the cases when the residual line-width can be interpreted in terms of AHT. At the same time, for $\Omega \approx |H|$ a significant line-narrowing can be observed in experiments. Therefore, the role of the higher-order terms remains unclear. If one assumes that the low-order approximation to the average Hamiltonian describes the true line shape, then the magnitude of this truncated Hamiltonian would be rather related to square root of the second moment of the line shape than the line width. The later two are very different for Lorentzian-like line shapes observed in the line-narrowing experiments. Alternatively, if the truncated average Hamiltonian contributes only to a central part of the spectral line (which, we think, is true in all practically interesting situations), the question of the line shape, width, and height remains open. In other words, after a low-order approximation to the average Hamiltonian or Liouvillian is calculated, it is not clear how to relate it to experimental line width or spectral peak height. In many applications, especially in multi-dimensional NMR experiments and also in the NMR investigation of solids, sensitivity is more important than resolution. Peak heights rather than line widths are commonly used to characterize efficiency of various line-narrowing techniques. In this letter, we show how a theory can be formulated to address calculation of the heights of spectral peaks.

For time-independent Hamiltonian $H$, the normalized free-induction decay (FID) signal $M(t)$ and the conventional NMR spectrum $I(\omega)$ can be written as

$$M(t) = Tr\{S_X \exp(-iHt) S_X \exp(iHt)\} / Tr\{S_X^2\} = Tr\{S_X \rho_t(t)\} / Tr\{S_X \rho_t(0)\}, \quad (1a)$$

$$I(\omega) = Re \int_0^\infty dt\, \exp(-i\omega t) M(t) = Re\, Tr\{S_X \rho_\omega(\omega)\} / Tr\{S_X \rho_t(0)\}, \quad (1b)$$

where $S_X$ is the x-component of the total spin angular momentum operator, $\rho_t$ and $\rho_\omega$ are the density matrices in time and frequency domains respectively. By introducing the Liouvillian superoperator $L = -i[H, \ldots]$, the binary product $\langle A | B \rangle = Tr(A^+ B)$, shorter notation $S_X = |x\rangle$, normalization $\langle x | x \rangle = 1$, and the initial density matrix $\rho_t(0) = |x\rangle$, one can rewrite Eq.(1) as

$$M(t) = \langle x | \rho_t(t) \rangle, \qquad |\rho_t(t)\rangle = \exp(Lt) |x\rangle, \quad (2a)$$

$$I(\omega) = Re \langle x | \rho_\omega(\omega) \rangle, \qquad |\rho_\omega(\omega)\rangle = \int_0^\infty dt\, \exp(-i\omega t) |\rho_t(t)\rangle = (i\omega - L)^{-1} |x\rangle. \quad (2b)$$

Some technical details on using super-operators and super-resolvents can be found in ref. [10]. For mathematical simplicity, in Eq.(2b) and below, we assume that the Liouvillian contains small damping $\varepsilon$: $L' = L - \varepsilon$, which is set to zero at the end of all calculations by taking the limit $\varepsilon \to 0$. Expansions of Eqs. (2a) and (2b) in powers of $L$ are

$$M(t) = \sum_{n=0}^\infty (t^n/n!) \langle x | L^n | x \rangle \quad (3a)$$

$$I(\omega) = Re\, (1/i\omega) \sum_{n=0}^\infty \langle x | (L/i\omega)^n | x \rangle. \quad (3b)$$

Eq. (3a) is the familiar moment expansion of FID [11]. Eq. (3b) can be obtained from (2b) by iterative application of the identity

$$(A - B)^{-1} = A^{-1} + (A - B)^{-1} B A^{-1}. \quad (4)$$

The first term of expansion (3b) is $Re\,(1/i\omega) = \delta(\omega)$, according to our convention of using small damping $\varepsilon$.

Now we will derive Eq. (2b) in a way, suitable for time-dependent Liouvillian $L(t)$. In particular, we assume that the Hamiltonian and, therefore, the Liouvillian, are periodic functions of time with the period $2\pi/\Omega$:

$$L(t) = \Sigma_n L_n \exp(in\Omega t). \quad (5)$$

The equation of motion

$$(d/dt) |\rho_t(t)\rangle = L(t) |\rho_t(t)\rangle = \Sigma_n L_n \exp(in\Omega t) |\rho_t(t)\rangle \quad (6)$$

after a Fourier transform becomes

$$\int_0^\infty dt\, \exp(-i\omega t)(d/dt) |\rho_t(t)\rangle = \int_0^\infty dt\, \exp(-i\omega t) \Sigma_n L_n \exp(in\Omega t) |\rho_t(t)\rangle,$$

$$i\omega |\rho_\omega(\omega)\rangle - |x\rangle = \Sigma_n L_n |\rho_\omega(\omega - n\Omega)\rangle. \quad (7)$$

By introducing the operator of frequency shift $T$ as $T |\rho_\omega(\omega)\rangle = |\rho_\omega(\omega - \Omega)\rangle$, one can rewrite Eq.(7) in a more compact form as

$$(i\omega - L) |\rho_\omega(\omega)\rangle = |x\rangle, \quad \text{or} \quad |\rho_\omega(\omega)\rangle = (i\omega - L)^{-1} |x\rangle, \quad (8)$$

where

$$L = \Sigma_n L_n T^n. \tag{9}$$

With the new Liouvillian (9), Eq.(8) has exactly the same form as Eq.(2b) derived for the case of time-independent Liouvillian. At $\Omega \to 0$, $T \to 1$, $L = \Sigma_n L_n T^n \to \Sigma_n L_n = L(0)$, and Eqs.(8) and (2b) coincide. Similar to Eq.(3), expansion of the second equation of Eqs.(8) in powers of $L$ is not suitable for exploring the limit $\omega \to 0$. Instead, we will use the first of Eqs.(8), without inversion, as a set of coupled equations for $|\rho_\omega(\omega)>$, $|\rho_\omega(\omega + \Omega)>$, $|\rho_\omega(\omega - \Omega)>$, ... , and introduce shorter notations $|\rho_\omega(n\Omega)> = |\rho_n>$. The quantity of interest will be $I(0) = Re <x|\rho_\omega(0)> = Re <x|\rho_0>$.

In NMR experiments involving line-narrowing by coherent averaging, like homo- or hetero-nuclear dipolar decoupling or magic-angle spinning (MAS), at $\Omega \approx |H|$ the intensity of the central band is already much higher than that of the satellites or spinning side bands: $I(0) \gg I(\pm\Omega) \gg I(\pm 2\Omega) \gg \ldots$ . In this case, neglecting the terms $|\rho_n>$ with $|n| > 1$ would be a reasonable approximation. With this truncation and $L_0 = 0$, the set of equations (8) for $|\rho_n>$ becomes

$$\begin{aligned} -L_1|\rho_{-1}> - L_{-1}|\rho_1> &= |x> \\ i\Omega|\rho_1> - L_1|\rho_0> &= |x> \\ -i\Omega|\rho_{-1}> - L_{-1}|\rho_0> &= |x>. \end{aligned} \tag{10}$$

The solution of Eqs.(10) for $|\rho_0>$ is

$$|\rho_0> = [L_1, L_{-1}]^{-1}(i\Omega - L_1 + L_{-1})|x>. \tag{11}$$

Therefore, the central peak height $I(0) = Re <x|\rho_0>$ is a sum of linear in $\Omega$ term, described by the "average Liouvillian" $-[L_1, L_{-1}]/i\Omega$, and the term which does not depend on $\Omega$. The conventional second-order average Liouvillian is

$$L_{AV}^{(2)} = (\Omega/2\pi)\int_0^{2\pi/\Omega} dt' \int_0^{t'} dt'' L(t')L(t'') = -\sum_{n>0} [L_n, L_{-n}]/(in\Omega). \tag{12}$$

It can be also obtained by using the Liouvillian in Eq.(9) in the expansion in Eq.(3b). Compared to Eq.(12), Eq. (11) contains only the term with $n = 1$. More important difference is in the way how this Liouvillian is used. The inverse "average Liouvillian" in Eq.(11) is directly related to experimentally observable height of the spectral peak.

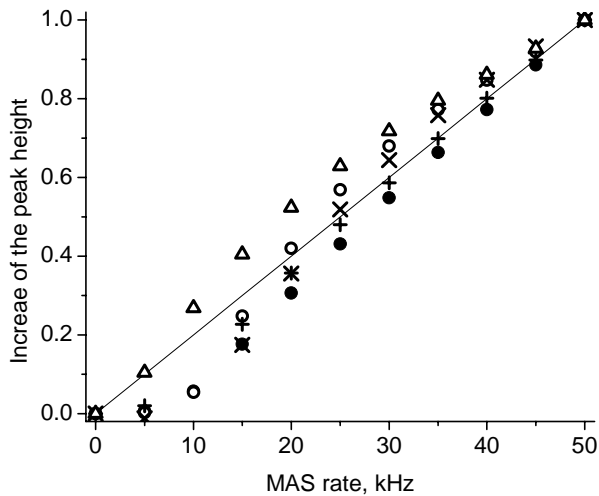

FIG. 1. Dependence of the experimentally measured increase in the $^1$H NMR peak height on MAS rate in powder samples of : (●) glycine, (○) glucose, (+) valine-leucine dipeptide line 1 ($CH_3$), (x) valine-leucine dipeptide line 2, (Δ) adamantane. All experiments were carried out on 600 MHz Varian solid-state NMR spectrometer using a 1 mm ultrafast-spinning double-resonance MAS probe. A single-pulse excitation with a recycle delay of 5 s and 4 scans were used.

Fig.1 shows the experimentally measured increase in $^1$H NMR peak heights for several powder samples as a function of MAS rate. The heights are normalized so that the increase measured at 50 kHz MAS rate is assigned to be one. It should be noted that the data represent a very broad range of actual peak heights, exceeding the peak height in static samples by one to two orders of magnitude. For the valine-leucine dipeptide, there are two well-resolved peaks at 50 kHz MAS. The height of the line at fixed frequencies of these peaks was measured at all spinning rates for this sample. The results for all samples except adamantane look similar. In adamantane, intra-molecular dipolar couplings are averaged out by fast molecular rotations, resulting in 13 kHz residual static line width, so that 5 kHz spinning causes significant line-narrowing, and we already expect a linear dependence on Ω. There should be a decrease in slopes for all samples at very high spinning rates, because other line-broadening factors, including omitted contributions to $L_0$ (distribution of isotropic chemical shifts, imperfect shims and any inaccuracy in the magic angle set-up) and spin-lattice relaxation, were not taken into account. Ref. [12] presents experimental dependencies of line widths on MAS rate for a variety of samples. In most cases, the dependences for line width are much more irregular than the data presented in Fig. 1.

In conclusion, expansions in powers of the interaction are not suitable for describing heights of spectral peaks. On the other hand, as demonstrated in this study, expansions in powers of the inverse interaction are valuable in the analysis of peak heights. While for time-independent Hamiltonian the limit ω → 0 in Eq.(2b) does not provide any new insights, for periodically modulated Hamiltonians it can lead to meaningful approximations like in Eq. (11).

AKK was a visiting Professor at the University of Michigan during his sabbatical leave. This study was supported by research funds from NIH (GM084018, GM095640 and RR023597to A.R.) and the instrumentation grant from CRIF-NSF.